\documentclass[preprint, showpacs, aps,draft]{revtex4}
\usepackage{bm}

\begin{document}
\title{Comment on ``An Arrow of Time Operator for Standard Quantum Mechanics'' (a sign of the time!)}
\author{Michael J. W. Hall}
\affiliation{Theoretical Physics, IAS, \\ Australian National
University,\\
Canberra ACT 0200, Australia}

\begin{abstract}
It is shown that the `arrow of time' operator, $\hat{M}_F$, recently suggested by Strauss et al., in arXiv:0802.2448v1 [quant-ph], is simply related to the sign of the canonical `time' observable, $T$ (apparently first introduced by Holevo).  In particular, the monotonic decrease of $\langle \hat{M}_F\rangle$ corresponds to the fact that $\langle \,{\rm  sgn~}T\rangle$ increases monotonically with time.  This relationship also provides a physical interpretation of the property $\hat{M}_F\leq \hat{1}$.  Some further properties and generalisations are pointed out, including to almost-periodic systems.
\end{abstract}

\maketitle
 
\section{Introduction}

Strauss et al. have recently given an interesting example of a `Lyapunov operator' applicable to a wide class of quantum systems \cite{strauss}, i.e., an operator which has a monotonically decreasing expectation value for all initial states.  In particular, for any system with a Hamiltonian operator $\hat{H}$ with eigenstates of the form 
\begin{equation} \label{hform}  
\hat{H}|E,j\rangle = E |E,j\rangle,~~~~~E\in [0,\infty),~~~j=1,2,\dots,d 
\end{equation}
for some fixed $d$ that is independent of $E$, Strauss et al. define the operator 
\begin{equation} \label{mf} 
\hat{M}_F := \frac{i}{2\pi}\sum_j \int_0^\infty dE \int_0^\infty dE'\, \frac{|E,j\rangle\,\langle E',j|}{ E-E' +i0^+} ,  
\end{equation}
and demonstrate that for any state $|\psi_t\rangle$ the expectation value 
$\langle \hat{M}_F\rangle_{\psi_t}$ 
 is monotonic decreasing with time \cite{strauss}.  Strauss et al. further give some numerical examples, and determine the eigenstates of $\hat{M}_F$.

The physical origin of $\hat{M}_F$ above is not particularly obvious, and earlier work of Strauss only provides a rather mathematical motivation, related to a Hardy space representation of the Schr\"{o}dinger equation \cite{s1,s2}. Here it will be shown that the operator $\hat{M}_F$ is in fact closely related to the canonical time observable, $T$, apparently first introduced by Holevo \cite{holevo}.  In particular, one has the general relation
\begin{equation} \label{sign}  
\langle \hat{M}_F \rangle \equiv \frac{1}{2}\left( 1- \langle \,{\rm sgn}~ T \rangle \right).  
\end{equation}
Hence, {\it the Lyapunov operator is closely related to the sign of the canonical time observable}, providing a simple physical interpretation for the former.

It is important to note that the canonical time observable $T$ is a probability operator measure (POM) \cite{holevo}, and hence is described by a set of positive operators which sum to the identity operator \cite{holevo,nielsen}, i.e.,
\[ T \equiv \{ \hat{T}_t \}, ~~~~~\hat{T}_t\geq0,~~~~~~\int_{-\infty}^\infty dt\,\hat{T}_t = \hat{1} , \]
with the probability density for a measurement of $T$ to give result $t$ for state $|\psi\rangle$ given by
\begin{equation} \label{pt} 
p_T(t|\psi) = \langle \psi|\hat{T}_t|\psi\rangle  . 
\end{equation}
Such POM observables are well known to be essential for describing all possible measurements that may be made on a given quantum system, and may always be represented in terms of measurement of a Hermitian operator on an `apparatus' system which has interacted with the system \cite{holevo, nielsen}.  The main advantage of the POM formalism is that one does not have to describe such apparatus systems explicitly, when considering the possible measurements on a given system, which is particularly useful when determining optimal measurements for extracting information in various scenarios.

In the next section the origin and basic properties of the canonical time observable $T$ are reviewed. In particular,  $T$ is the optimal observable for covariantly estimating time translations, i.e., it is the optimal `clock' observable for the system. The relation (\ref{sign}) between $\hat{M}_F$ and $T$ is demonstrated in section III, and the corresponding physical interpretation of $\hat{M}_F$ is discussed.  Some further properties and possible generalisations are discussed in Sec.~IV.

\section{Canonical time observable}

For a quantum system with energy eigenstates as per equation (\ref{hform}), define the corresponding `time' kets 
\[  |t,j\rangle := (2\pi)^{-1/2} \int_0^\infty dE\, e^{-iEt} |E,j\rangle~~~~~~~j=1,2,\dots,d.  \]
Note that natural units with $\hbar=1$ have been adopted, in keeping with \cite{strauss}.  The corresponding canonical time observable $T$ is then defined as the POM observable $\{\hat{T}_t\}$, with
\begin{equation} \label{tt}  
\hat{T_t} := \sum_j |t,j\rangle\,\langle t,j| .  
\end{equation}
It is easily checked that 
\[ \int_{-\infty}^\infty dt\,\hat{T}_t = \sum_j \int_0^{\infty} dE\,|E,j\rangle\,\langle E,j| = \hat{1} , \]
as required.  This `canonical' time observable appears to have first been considered in some detail by Holevo, primarily for the case of a free particle \cite{holevo}.  

It is worth noting some properties of $T$ here, to indicate why it is a natural time observable to consider at all.  First, observe that if $E$ and $t$ were replaced by momentum and position coordinates $p$ and $x$ in the above definition of $|t,j\rangle$, with the range of integration extended over the whole real line, then one would obtain the usual Fourier relation between conjugate position and momentum kets.  Hence, by analogy, $T$ can be said to be conjugate to the energy observable $\hat{H}$.  Indeed, the (truncated) Fourier relation between $|E,d\rangle$ and $|t,d\rangle$ immediately implies the entropic uncertainty relation 
\begin{equation} \label{ent} L_H L_T \geq  \pi e\hbar  , 
\end{equation}
precisely as for the case of position and momentum observables \cite{bbm}, where the `ensemble length' $L_A$ is the natural geometric measure of the spread of observable $A$, given by the exponential of the entropy of $A$ \cite{ensemble}.  Hence, the energy and time uncertainties cannot simultaneously be arbitrarily small.  

Second, it follows immediately from Eqs.~(\ref{pt}) and (\ref{tt}) that
\begin{equation} \label{cov}
p_T(t'|\psi_t) = p_T(t'-t|\psi_0)  ,
\end{equation}
i.e., the probability distribution simply translates under time evolution of the system.  This time-covariance property is of course expected of any good `clock' observable \cite{holevo}.  Note in particular it implies that
\begin{equation} \label{clock}  
\langle T\rangle_{t} := \int_{-\infty}^{\infty}dt'\,t'\,p_T(t'|\psi_t) = \langle T\rangle_{0} + t  . 
\end{equation}

Third, Holevo has shown that the canonical time observable provides the best estimate of an unknown time shift of the system, for a particular figure of merit \cite{holevo}.  It may further be shown that $T$ is optimal in the sense that measurement of any {\it other} time-covariant observable is equivalent to first subjecting the system to some `noise' process, and then making a measurement of $T$ - see, for example, the analogous property in Ref.~\cite{hallqcm} for optical phase.

Finally, it should be remarked that the `time' kets $|t,j\rangle$ are {\it not} mutually orthogonal, due to the semiboundedness of the energy spectrum.  Hence, $T$ cannot correspond to some Hermitian operator on the Hilbert space of the system. Indeed, the truncated Fourier transform defining $|t,j\rangle$ implies, via the Paley-Wiener theorem, that $p_T(t|\psi)$ cannot vanish on any non-zero finite interval \cite{paley}, and so the canonical time distribution is always `fuzzy', with support over the entire real axis. Of course, for each real function $f(t)$, the corresponding average value of $f(T)$ follows  from (\ref{pt}) as
\begin{equation} \label{ft} 
\langle f(T)\rangle_\psi = \int_{-\infty}^{\infty}dt\,f(t)\,p_T(t|\psi) = \int_{-\infty}^{\infty}dt\,f(t)\, \langle \psi|\hat{T}_t|\psi\rangle, 
\end{equation}
and hence one can define a corresponding Hermitian operator 
\begin{equation} \label{ftop} \widehat{f(T)} := \int_{-\infty}^\infty dt\, f(t)\hat{T}_t  
\end{equation}
satisfying 
\[ \langle f(T)\rangle_\psi = \langle\psi |\widehat{f(T)}|\psi\rangle   \]
for all states $\psi$.
In general, however, these operators are not simply related algebraically - for example, one does not have $\widehat{(T^2)}=(\widehat{T})^2$.  Hence, it is the POM $T$ which is of fundamental significance, rather than any particular Hermitian operator $\widehat{f(T)}$.  This has a bearing on the interpretation of the Lyapunov operator in Eq.~(\ref{mf}), which corresponds to a particular choice of $f(t)$.

\section{`Arrow of time'  vs `canonical' time}

Consider now the observable corresponding to the {\it sign} of the canonical time observable, where ${\rm sgn~}t$ is defined to be $-1$, $0$ and $+1$ for $t<0$, $t=0$ and $t>0$ respectively.  By definition this observable can only have measured values in $\{-1, 0, 1\}$, and hence one must have
\begin{equation} \label{bound}  -1 \leq \langle\,{\rm sgn~}T \rangle_\psi \leq 1 .  
\end{equation}
Moreover, recalling from Eq.~(\ref{cov}) that the probability distribution of $T$ moves to the `right' as $t$ increases, one expects that the sign of $T$ must increase monotonically on average, just as $T$ itself does as per Eq.~(\ref{clock}).  Indeed, noting that, trivially, ${\rm sgn~}(t'+t)\geq {\rm sgn~}t'$ for $t\geq 0$, one has from Eq.~(\ref{cov}) that
\begin{equation} \label{mono} 
\langle\,{\rm sgn~}T \rangle_{\psi_t} = \int dt' \,{\rm sgn~}t'\, p_T(t'-t)|\psi_0) = \int dt' \,{\rm sgn~}(t'+t)\, p_T(t'|\psi_0) \geq \langle \,{\rm sgn~}T \rangle_{\psi_0}  . 
\end{equation}
In fact, strict inequality holds for $t>0$, since $p_T(t|\psi)$ cannot vanish on any non-zero finite interval as noted in the previous section.

It is seen from Eqs.~(\ref{clock}) and (\ref{mono}) that both $T$ and ${\rm sgn~}T$ have monotonic increasing expectation values.  It then follows immediately that, for example, $-T$ and $-{\rm sgn~}T$ both have monotonic {\it decreasing} expectation values for all initial states.  Hence, in particular, the corresponding Hermitian operators $-\widehat{T}$ and $-\widehat{{\rm sgn~}T}$ defined via Eq.~(\ref{ftop}) are `Lyapunov' operators in the sense of Strauss et al. \cite{strauss}.  Clearly there are many more such operators, but what is of interest here is the connection between $T$ and the particular Lyapunov operator $\hat{M}_F$ in Eq.~(\ref{mf}).

In particular, from Eqs.~(\ref{tt}) and (\ref{ftop}), and using a standard table of Fourier transforms, one has the explicit expression
\begin{eqnarray*}
 \widehat{{\rm sgn~}T } & = & (2\pi)^{-1}\sum_j  \int dE\int dE' |E,j\rangle\langle E',j| \int dt\,({\rm sgn~}t)\,e^{-i(E-E')t}\\
 & = & (2\pi)^{-1} \sum_j  \int dE\int dE' |E,j\rangle\langle E',j|~~ {\rm P.V.}\left[ -2i(E-E')^{-1}\right] \\
& = & 1 - 2 \hat{M}_F  ,
\end{eqnarray*}
where P.V. denotes the principal value, and the last line follows via the definition of $\hat{M}_F$ in Eq.~(\ref{mf}).  One hence obtains the desired simple relation in Eq.~(\ref{sign}) between $\hat{M}_F$ and ${\rm sgn~}T$.

Note that the property $0\leq \hat{M}_F\leq 1$, and the monotonic decrease of the expectation value of $\hat{M}_F$ with time, both proved by Strauss et al. \cite{strauss}, follow immediately from Eqs.~(\ref{sign}), (\ref{bound}) and (\ref{mono}).  They are seen to correspond to (i) the property $|\,{\rm sgn~}t|\leq 1$ and (ii) the monotonicity of ${\rm sgn~}t$.  Further, the {\it strict} monotonic decrease of $\langle \hat{M}_F \rangle$ follows from the property that the canonical time distribution has support over the whole real axis (up to a set of measure zero).  However, the eigenfunctions of $\hat{M}_F$, determined by Strauss et al. \cite{strauss}, are seen not to have any particular fundamental significance - they are merely eigenfunctions of an operator corresonding to a particular function of the canonical time observable, where different eigenfunctions would be obtained by choosing a different function of $T$.

\section{Discussion}

It has been shown that the particular Lyapunov operator $\hat{M}_F$, investigated by Strauss et al. \cite{strauss}, has a simple relationship to the sign of the canonical time observable $T$, as per Eq.~(\ref{sign}), and that its main properties can be easily obtained from general properties of $T$.  Further, many other Lyapunov operators can be constructed from $T$ - in particular, the operator $\widehat{g(T)}=\int dt\,g(t)\,\hat{T}_t$, for any monotonic decreasing function $g(t)$.

It is worth noting that in practice it is simplest to determine $\langle\hat{M}_F\rangle$, for a given state $|\psi\rangle$, by first calculating 
\[ p_T(t|\psi_0) = \sum_j \left| \langle t,j|\psi_0\rangle \right|^2 , \]
and then using Eqs.~(\ref{sign}) and (\ref{cov}) to calculate 
\begin{equation} \label{simp} 
\langle \hat{M}_F\rangle_{\psi_t} =  \int_{-\infty}^0 dt'\,p_T(t'-t|\psi_0)= \int_{-\infty}^{-t} dt'\,p_T(t'|\psi_0) . 
\end{equation}
Note it follows that $\langle M_F\rangle$ is the cumulative probability distribution for the canonical time observable, thus providing a physical interpretation for Eq.~(7) in Ref.~\cite{strauss}.

For example, for an initially stationary free particle of mass $m$ in one dimension, with Gaussian momentum representation
\[ \psi_0(p) =  e^{-p^2/(4\sigma^2)} \]
up to a normalisation factor, one  finds the Gaussian integral
\[ \langle t,\pm|\psi_0\rangle = \int_0^\infty dp\,p^{1/2} e^{ip^2t/(2m)} e^{-p^2/(4\sigma^2)} \]
up to a normalisation factor (where $p=\pm (2mE)^{1/2}$), thus yielding a distribution of the form
\[ p_T(t|\psi_0) = N \left[ t^2 + (m/2\sigma^2)^2\right]^{-3/2}  \]
for the canonical time observable, for some normalisation constant $N$. One therefore obtains from Eq.~(\ref{simp}), restoring general units, the explicit result
\[ \langle \hat{M}_F\rangle_{\psi_t}= 1/2 - (1/2)\,t\, [t^2 + (\hbar m/2\sigma^2)^2]^{-1/2}  . \]

Note that the case of a free particle does not quite satisfy the assumed spectral condition in Eq.~(\ref{hform}), as the degeneracy breaks down at $E=0$.  While not affecting the above results, this property leads, for example, to a divergence in the variance of the canonical time observable when the wavefunction has a non-zero component corresponding to $E=0$ \cite{holevo}, which can be seen to occur for the free-particle example above.  Hence, no Heisenberg type uncertainty relation can be written down in this case.  In contrast, the ensemble lengths $L_H$ and $L_T$ in the time-energy uncertainty relation (\ref{ent}) are perfectly well-defined for the above example \cite{ensemble}, and can be explicitly calculated using standard tables of integrals (allowing evaluation of how close their product is to the lower bound $\pi e\hbar$).

It is of interest to consider how the canonical time observable may be generalised to quantum systems with energy spectra different to that of Eq.~(\ref{hform}).  This is quite simple for an evenly spaced discrete energy spectrum, such as a harmonic oscillator, where essentially the Fourier transform defining the `time' kets $|t,j\rangle$ is replaced by a discrete Fourier transform, yielding a periodic canonical time observable \cite{holevo,helstrom,hallqcm}.  This replacement also applies to any energy spectra which is a subset of an evenly spaced discrete set.  More generally, for an arbitrary continuous or uniform discrete energy spectrum, with a possibly continuous degeneracy which may depend on $E$, one may formally extend the energy spectrum to the form $\{|E,j\rangle\}$ with $j$ ranging over some sufficiently large measurable set $J$, with $|E,j\rangle=0$ for some values of $j$ if the degeneracy is not uniform, and define the corresponding canonical time observable $T\equiv \{\hat{T}_t\}$ as per Eq.~(\ref{tt}), where the $|t,j\rangle$ are defined as before (with integration replaced by summation for discrete energy spectra).  

The cases of mixed continous and discrete energy spectra, and of discrete non-uniform energy spectra, appear to be more difficult - such systems simply may not make good `clocks' (certainly this would be the case for chaotic classical systems).  Note, however, that if the state of the system only has support on the continuous portion of the spectrum, or on a uniformly spaced subset of the discrete portion, then a time observable may be defined as above on the corresponding restricted Hilbert space.  

Finally, it is of interest to note, assuming uniform degeneracies for convenience, that for a general discrete energy spectrum $\{ |E_k\rangle\} $ one may be able to at least speak of `clocks' relative to a certain resolution.  In particular, one can certainly always define a POM $T(\tau):=\{ \hat{T}_t(\tau);\hat{P}(\tau)\}$, with $t$ taking values in $[0,\tau)$, by
\begin{equation} 
\hat{T}_t(\tau):= \sum_j \hat{N}_\tau^{-1/2} |t,j\rangle\langle t,j| \hat{N}_\tau^{-1/2}, 
\end{equation}
where $|t,j\rangle$ is defined as in Sec.~II (with integration replaced by summation), $N_\tau$ is the positive operator
\[ \hat{N}_\tau := \sum_j \int_{0}^\tau dt\,|t,j\rangle\langle t,j|  , \]
$\hat{N}_\tau^{-1/2}$ is defined to be $0$ when acting outside the support of $N_\tau$, and $\hat{P}_\tau$ is the projection operator onto the zero eigenspace of $\hat{N}_\tau$.  The limit $\tau\rightarrow\infty$ is well defined when the energy spectrum is uniform spaced, yielding a periodic time observable.  More generally, the evolution will be almost-periodic, and for a given resolution $\epsilon$ there will be a period $\tau_\epsilon$ for which the system state will evolve arbitrarily close to its (arbitrary) initial state, to within a distance defined by $\epsilon$.  Hence, it appears that one may define a time observable for such systems relative to a given resolution parameter $\epsilon$.  This would be of interest for further investigation.

\end{document}